\documentclass[twocolumn]{jpsj2} 
\newcommand{\GaPhNo}{$\beta''$-(BEDT-TTF)$_{4}$[(H$_{3}$O)Ga(C$_{2}$O$_{4}$)$_{3}]\cdot$C$_{6}$H$_{5}$NO$_{2}$}

\title{$^{13}$C NMR Study of Superconductivity Near Charge Instability Realized in \GaPhNo}
\author{Yoshihiko~Ihara,
Harumi~Seki, and Atsushi~Kawamoto}

\inst{Department of Physics, Graduate School of Science, Hokkaido University, Sapporo 060-0810, Japan.}
\date{\today}

\abst{
To investigate the superconducting (SC) state near a charge instability,
we performed $^{13}$C NMR experiments on the molecular superconductor \GaPhNo~ (BEDT-TTF: bis(ethylenedithio)tetrathiafulvalene), 
in which charge disproportionation was reported from Raman spectroscopy at $100$ K. 
We found an NMR spectral splitting at $T_{\rm CD}=12$ K, which is ascribed to the low-temperature charge instability. 
Measurements of nuclear spin-lattice relaxation time revealed enhanced fluctuations at $T_{\rm CD}$. 
We suggest the relationship between the enhanced fluctuations and superconductivity, 
which emerges slightly below $T_{\rm CD}$.   
In the SC state, we observed a decrease in the Knight shift, which leads us to suggest a spin-singlet SC state. 
The possibility of an unconventional SC state is discussed. 
}

\kword{organic superconductor, charge order, NMR}

\begin{document}

\maketitle

Superconductivity appearing in close proximity to magnetism has fascinated researchers 
because of its unconventional superconducting (SC) pairing mechanisms. 
For example, in the Ce-based heavy-fermion superconductor CeCu$_{2}$Si$_{2}$, 
Cooper pairs are mediated by magnetic fluctuations 
that are enhanced near the magnetic quantum critical point (QCP) \cite{steglich-PRL43, mathur-nature394}.
A similar scenario has been applied to the interpretation of the SC state in a high-SC-transition-temperature cuprate, pnictide, 
as well as in organic superconductors such as (TMTSF)$_{2}$PF$_{6}$ \cite{jerome-physique41} (TMTSF: tetramethyltetraselenafulvalene) 
and $\kappa$-(BEDT-TTF)$_{2}X$ \cite{williams-science252, murakami-JPSJ69} (BEDT-TTF: bis(ethylenedithio)tetrathiafulvalene), 
and it has successfully explained the physical properties. 
Holmes {\it et al.} \cite{holmes-PRB69} have suggested a novel mechanism 
that is triggered by an increase in the SC transition temperature $T_{c}$ of CeCu$_{2}$Si$_{2}$ 
at pressures greater than $3$ GPa, 
at which the system approaches the second critical point ascribed to a valence instability. 
Theory indicates that enhanced charge fluctuations near an \emph{electric} QCP increase $T_{c}$.  
Thus an experimental study that investigates superconductivity near charge instability is required.

Charge ordering is observed in several BEDT-TTF salts with $\alpha$- and $\theta$-type structures \cite{miyagawa-PRB62, takano-SM120}, 
some of which undergo SC transition in the charge-ordered state \cite{nishikawa-PRB72, morinaka-PRB80}. 
A theoretical study shows that 
an unconventional SC state can be realized in these compounds near charge instability \cite{merino-PRL87, watanabe-JPSJ74, merino-PRL96}, 
as expected for other superconductors observed near magnetic QCPs. 
Since optical experiments in BEDT-TTF salts have clearly revealed charge instability 
and indicated the connection between superconductivity and charge fluctuation \cite{tanaka-JPSJ77, kaiser-PRL105}, 
investigations of electronic properties in both the normal and  SC states by various experimental techniques are essential.

Superconductivity was observed at rather high temperatures reaching $9$ K in the $\beta ''$-type structures 
$\beta''$-(BEDT-TTF)$_{4}$[(H$_{3}$O)$M$(C$_{2}$O$_{4}$)$_{3}]\cdot Y$ with $M=$ Ga, Fe, or Cr and 
$Y=$ C$_{6}$H$_{5}$NO$_{2}$ or C$_{6}$H$_{5}$CN \cite{kurmoo-JACS117, akutsu-JACS124, coldea-PRB69, bangura-PRB72}. 
These compounds consist of alternating BEDT-TTF and anion block layers. 
The metallic conductivity is governed by the positive carriers injected into the two-dimensional BEDT-TTF layers. 
Magnetic anomaly is absent in these $\beta''$-family compounds, 
whereas the anomaly associated with charge ordering has been observed at approximately $100$ K by 
Raman spectroscopy and electric transport measurements \cite{bangura-PRB72}. 
Superconductivity in these salts must be investigated to determine the nature of superconductivity 
in the vicinity of charge instability. 

In addition to the interplay between charge ordering and superconductivity, 
the extremely high upper critical field $H_{c2}$ of $33$ T, which is almost three times the Pauli-Clogston limit, 
piques our interest \cite{bangura-PRB72}. 
In a magnetic field, superconductivity is suppressed by the Pauli depairing effect \cite{clogston-PRL9}. 
To sustain superconductivity in high magnetic fields, 
an unconventional SC state was introduced by Fulde and Ferrel and Larkin and Ovchinnikov (hereafter the FFLO state) \cite{fulde-PRA135, larkin-sov20}. 
In the FFLO state, the Pauli depairing effect is suppressed by allowing the real-space modulation of the SC gap. 
Whereas for the spin-triplet SC state, the Pauli depairing is irrelevant because 
Cooper pairs can preserve their spin degrees of freedom. 
To comprehend the high-field SC state, SC pairing symmetry must be determined from low-field experiments. 
NMR spectroscopy is a powerful technique for microscopically investigating the spin symmetry of Cooper pairs.
We have carried out the Knight shift ($K$) measurement in the SC state to determine the SC symmetry.

\begin{figure}[tbp]
\begin{center}
\includegraphics[width=8cm]{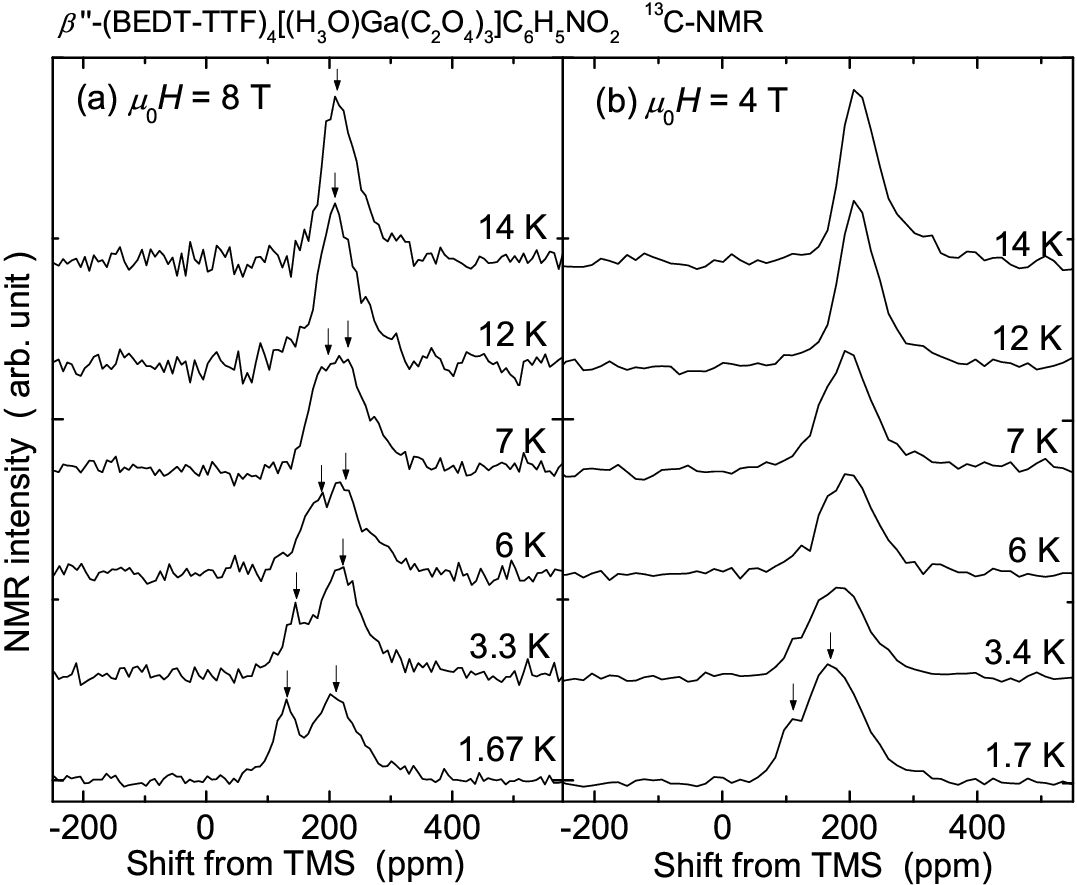}
\end{center}
\caption{
$^{13}$C NMR spectra of \GaPhNo\; at magnetic fields of (a) $8$ T and (b) $4$ T along the $b$ axis.
The NMR shift is measured from tetramethylsilane (TMS). 
The spectral splitting observed below $12$ K is ascribed to static charge instability.}
\label{fig1}
\end{figure}

In this Letter, we focus on the $\beta''$-salt with $M=$ Ga and $Y=$ C$_{6}$H$_{5}$NO$_{2}$, 
which exhibits superconductivity at $7.5$ K. 
Single-crystal samples of this salt were 
grown using the electrocrystallization technique described in ref. [\cite{akutsu-JACS124}]. 
We obtained two single-crystal phases: plate and needle crystals. 
The crystalline parameters of the needle crystals are consistent with  
those reported for the SC samples \cite{akutsu-JACS124}. 
Magnetization and resistivity measurements of the needle crystals detected SC transition at $T_{c}=7.5$ K, 
in accordance with a previous report \cite{akutsu-JACS124}, 
and the SC fraction estimated from magnetization data is the same as that in ref. [\cite{akutsu-JACS124}].
The typical dimensions of the SC needle samples are $1\times 0.1\times 0.2$ mm$^{3}$.
The single-crystal X-ray diffraction revealed that the longest axis is parallel to the crystalline $a$-axis. 
For the NMR experiment reported in this Letter, external magnetic fields were applied along the $b$-axis, 
which is the second-longest axis of the crystal. 
To reduce spectral broadening due to sample misalignment, 
$^{13}$C NMR spectra were acquired for one single crystal. 	 
To measure the spin-lattice relaxation time $T_{1}$, 
which does not require high frequency resolution, 
we aligned 30 single crystals on a flat sample holder to improve the signal-to-noise ratio.
For the resistivity measurement, 
we utilized a sample synthesized by the same process as those for the NMR experiment, 
and applied magnetic fields along the $b$-axis. 

When neighboring $^{13}$C nuclear spins are magnetically coupled, 
the NMR shift of the $^{13}$C resonance cannot be correctly determined, 
because the nuclear spin-spin coupling splits the $^{13}$C NMR spectrum into two peaks (Pake doublet\cite{pake-ChemPhys16}). 
For BEDT-TTF molecules, the coupling of $^{13}$C nuclei at the central C=C bond 
is sufficiently strong to result in this Pake-doublet problem \cite{kawamoto-PRB52}.
Therefore, we eliminated the Pake doublet by the selective enrichment of one side of the central C=C bond with $^{13}$C 
using the cross-coupling method \cite{yamashita-synthmet133} between 
non-enriched ketone and $^{13}$C-enriched thio-ketone forms. 

$\beta''$-salts possess two nonequivalent BEDT-TTF molecules, 
each of which possesses two nonequivalent $^{13}$C sites at the central C=C bond. 
The four nonequivalent $^{13}$C sites should result in four NMR peaks 
in a field applied exactly along the glide plane ($||b$).
However, as shown in Fig.~\ref{fig1}, we observed a single-peak NMR spectrum at $14$ K, 
because the NMR spectrum broadening starting from $100$ K, 
where charge ordering was detected by Raman spectroscopy, led to the merging of the split peaks. 
The absence of resolved spectral splitting is due to the similarity in crystallographic sites between 
two nonequivalent molecules, which is evident from the crystal structure \cite{akutsu-JACS124}. 
We found a clear peak splitting at low temperatures, as indicated by downward arrows in Fig.~\ref{fig1}(a). 
The spectral splitting indicates static $^{13}$C-site doubling caused either by the
crystallographic symmetry breaking or by the increase in disproportionation between two nonequivalent molecules. 
Spectral broadening below $12$ K and splitting at $1.7$ K were also observed at $4$ T, 
although the broader linewidth gives rise to a broad spectrum with a shoulder structure, as indicated by arrows in Fig.~\ref{fig1}(b).
The double-peak structure is well resolved only in high fields, 
because the high NMR frequency ($f_0$) used in high-field experiments improves the shift resolution. 

\begin{figure}[tbp]
\begin{center}
\includegraphics[width=7cm]{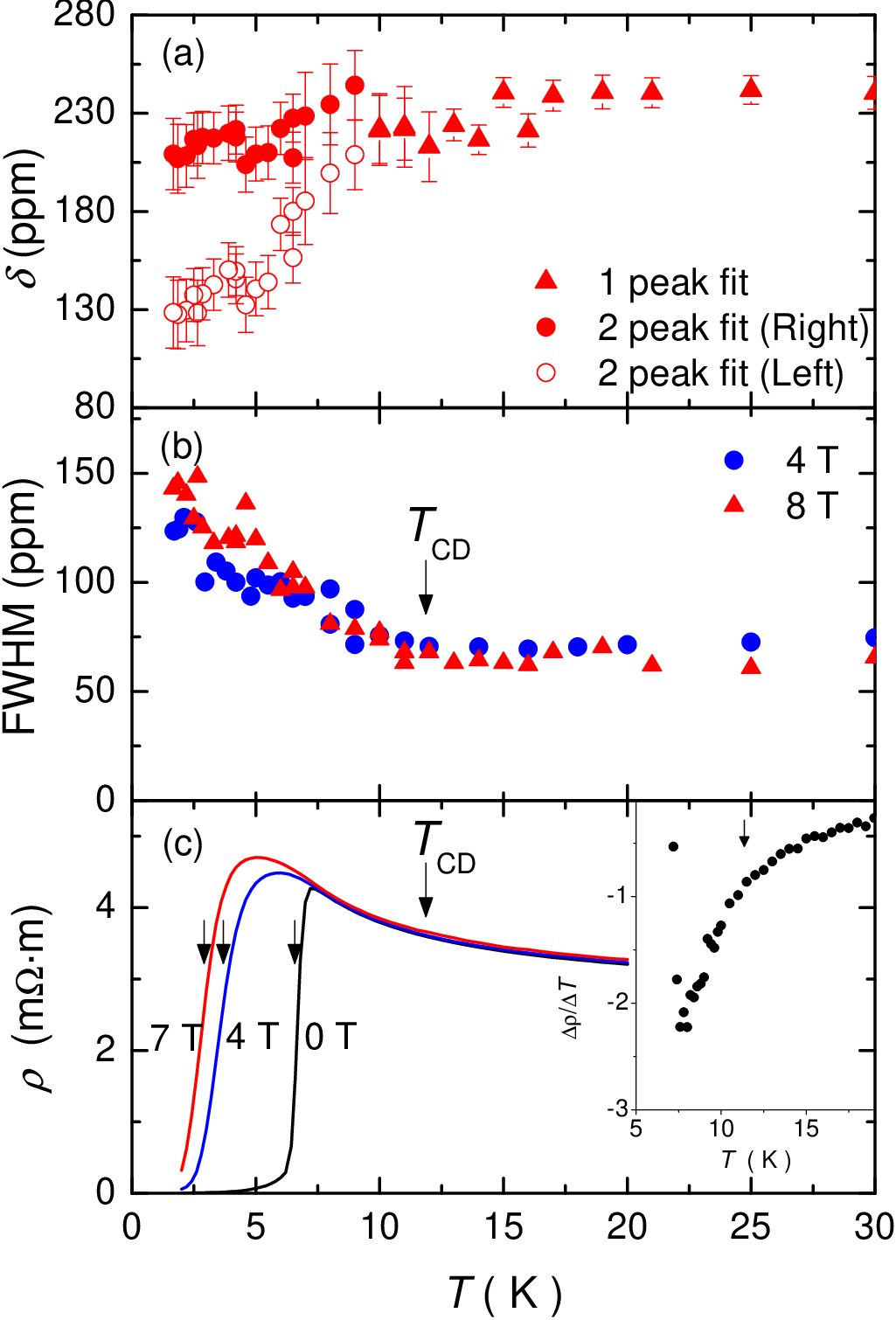}
\end{center}
\caption{(Color online)
(a) Temperature dependence of NMR shift in $8$ T. 
The peak positions were determined by fitting spectra with Lorentzians.
The results of the two-peak fit below $9$ K are plotted as solid and open circles. 
The spectrum becomes a single peak above $10$ K. 
The results of the single-peak fit above $10$ K are shown as solid triangles. 
(b) Full widths at half maximum (FWHMs) measured at $4$ and $8$ T. 
An abrupt increase was observed below $T_{\rm CD}=12$ K. 
The field independent FWHM indicates a charge instability for the anomaly at $T_{\rm CD}$. 
(c) In-plane resistivity measured at $H_{||b}$ of $0$, $4$, and $7$ T. 
$T_{c}$ is determined as $T_{c}$($0$ T) = $6.7$ K, $T_{c}$($4$ T) = $3.5$ K, and $T_{c}$($7$ T) = $2.8$ K 
from the midpoint.
The inset shows the temperature dependence of $\Delta \rho/\Delta T$. 
A change in slope was observed below $T_{\rm CD}$. 
}
\label{fig2}
\end{figure}

The temperature dependences of the NMR shift and full width at the half maximum (FWHM) are shown in Figs.~\ref{fig2}(a), \ref{fig2}(b), respectively. 
The spectral splitting was observed below approximately $9$ K, as shown in Fig.~\ref{fig2}(a). 
To determin the onset temperature, however, we adopted the FWHM data, 
because the splitting is detected as the spectral broadening 
when the separation of the two split peaks is smaller than the linewidth. 
Judging from the abrupt increase in FWHM,  
we determined the onset temperature of the spectral splitting to be $12$ K.  
At the same temperature, an anomaly was also found in resistivity. 
Figure \ref{fig2}(c) shows the in-plane resistivity measured at $0$, $4$, and $7$ T ($|| b$-axis). 
As shown in the inset of Fig.~\ref{fig2}(c), 
a gradual change in $\Delta \rho /\Delta T$ was observed at approximately $12$ K. 
This behavior is independent of the applied magnetic field. 
The similar temperature dependences of FWHM and resistivity indicate that 
the anomaly that induces spectral splitting does reflect electronic properties.

The field-dependent decrease in resistivity at low temperatures was assigned to SC transition.  
It is noteworthy that superconductivity sets in from a semiconducting state.  
We defined $T_{c}$ as the temperature at which resistivity becomes half of the normal-state values. 
The actual values are $T_{c}(7~\rm{T} )=2.8$ K, $T_{c}(4~\rm{T} )=3.5$ K, and $T_{c}(0~\rm{T} ) = 6.7$ K. 
As $T_{c}$ for the zero field is much lower than that for the onset of spectral splitting 
and the splitting is clearly observed even in high fields, 
SC transition is excluded as the origin of spectral splitting. 
We can also exclude the magnetic instability by the field-independent FWHM below $12$ K. 
When an internal magnetic field induced by magnetic transition splits the NMR spectrum, 
the frequency separation ($\Delta f$) is independent of the external field. 
Therefore, when we use the NMR shift ($\delta=\Delta f/f_0$) as the horizontal axis, 
the separation should decrease in high fields. 
While in the paramagnetic state, as the magnetization is proportional to the external field, 
the linewidth in NMR shift is independent of the external field. 
The field-independent FWHM shown in Fig.~\ref{fig2}(b) clearly indicates that 
the anomaly at $12$ K is not caused by the order in spin degrees of freedom. 
We note that the $\beta''$-salt demonstrates two independent charge instabilities at $100$ K and $T_{\rm CD}= 12$ K.
As the low-temperature anomaly occurs close to $T_{c}$, 
we suggest a relationship between the SC mechanism and the fluctuations near charge instability. 

\begin{figure}[tbp]
\begin{center}
\includegraphics[width=7cm]{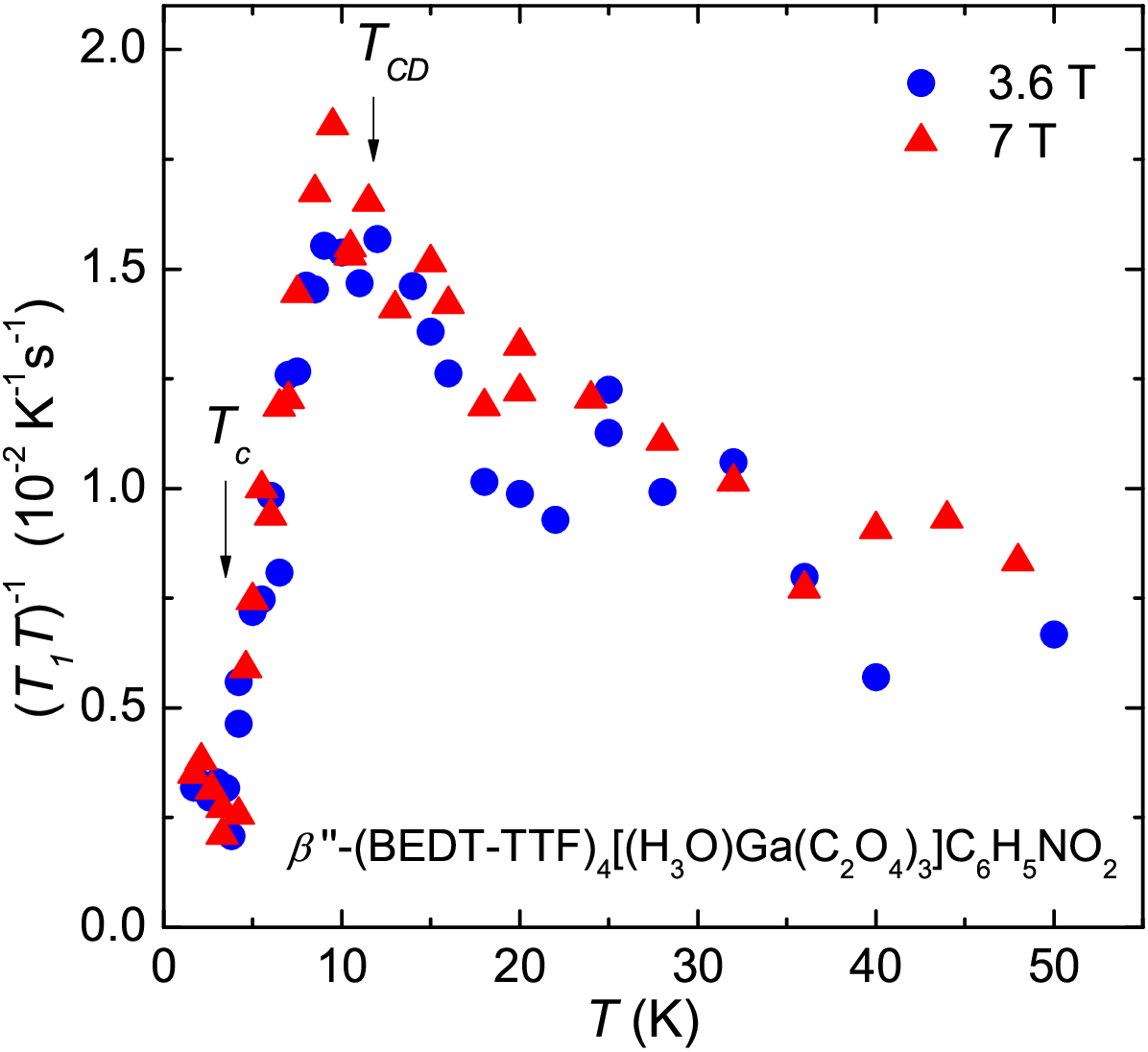}
\end{center}
\caption{(Color online)
Temperature dependences of $(T_{1}T)^{-1}$ measured at $3.6$ and $7$ T. 
The peak at $(T_{1}T)^{-1}$ was found at $T_{\rm CD}$ in both fields. 
The anomaly associated with SC transition was suppressed by the non-Fermi liquid behavior below $T_{\rm CD}$. 
}
\label{fig3}
\end{figure}

We investigated the dynamical properties above $T_{\rm CD}$ by measuring $T_{1}$ at $3.6$ T and $7$ T.
As shown in Fig.~\ref{fig3}, $(T_{1}T)^{-1}$ increases with decreasing temperature, forming a peak at $T_{\rm CD}$. 
In general, $(T_{1}T)^{-1}$ is expressed in terms of the dynamic susceptibility $\chi''(q,\omega)$ as
\begin{equation}
\frac{1}{T_{1}T} = \frac{2\gamma_n^2 k_B }{(\gamma_e \hbar )^2} \sum_q A_{q}A_{-q}\frac{\chi''(q,\omega)}{\omega}.
\end{equation}%
In the Fermi liquid state, $(T_{1}T)^{-1}$ is proportional to the square of the density of states, 
and is temperature-independent. 
The temperature dependence of $(T_{1}T)^{-1}$ is generated by the enhanced magnetic fluctuations in the vicinity of magnetic transition. 
However, for \GaPhNo\; with charge instability at $T_{\rm CD}$, 
magnetic fluctuations are so weak as they do not induce a strong temperature dependence. 
Charge fluctuations can be enhanced at $T_{\rm CD}$, 
but they cannot be directly detected by $^{13}$C NMR experiment, 
because $^{13}$C nuclei with a nuclear spin $I = 1/2$ do not have an electric quadrupole moment, 
which can interact with charge fluctuations. 
The coupling between charge and magnetic fluctuations is required 
to increase $(T_{1}T)^{-1}$ at $T_{\rm CD}$. 
One possible interpretation is that the fluctuations in local spin density, 
which are generated by charge density fluctuations, create fluctuating magnetic fields at the $^{13}$C site. 
Direct observation of charge fluctuations is desired to reveal the mechanisms of spin-charge coupling. 

In the charge-disproportionate state, where the NMR spectrum splits, 
$T_{1}$ was obtained using the integrated intensity of two peaks. 
We also measured $T_{1}$ using the right and left halves of the split spectrum, 
which results in the same value within experimental error. 
The uniform $T_{1}$ values over the entire NMR spectrum allow us to confirm that 
the spectrum splitting is not caused by macroscopic phase separation, 
but by the intrinsic instability of the electronic state. 
Below $T_{\rm CD}$, the Fermi liquid behavior in $(T_{1}T)^{-1}$ is absent until the SC state emerges.  
The non-Fermi liquid behavior is consistent with the semiconducting resistivity just above $T_{c}$. 
Since  $(T_{1}T)^{-1}$ decreases below $T_{\rm CD}$ following a power law close to $T^{2}$,  
a clear anomaly associated with SC transition was not observed at $T_{c}$. 
Note that $(T_{1}T)^{-1}$ keeps following a power-law behavior even in the SC state, 
which is suggestive of unconventional superconductivity.

\begin{figure}[tbp]
\begin{center}
\includegraphics[width=7cm]{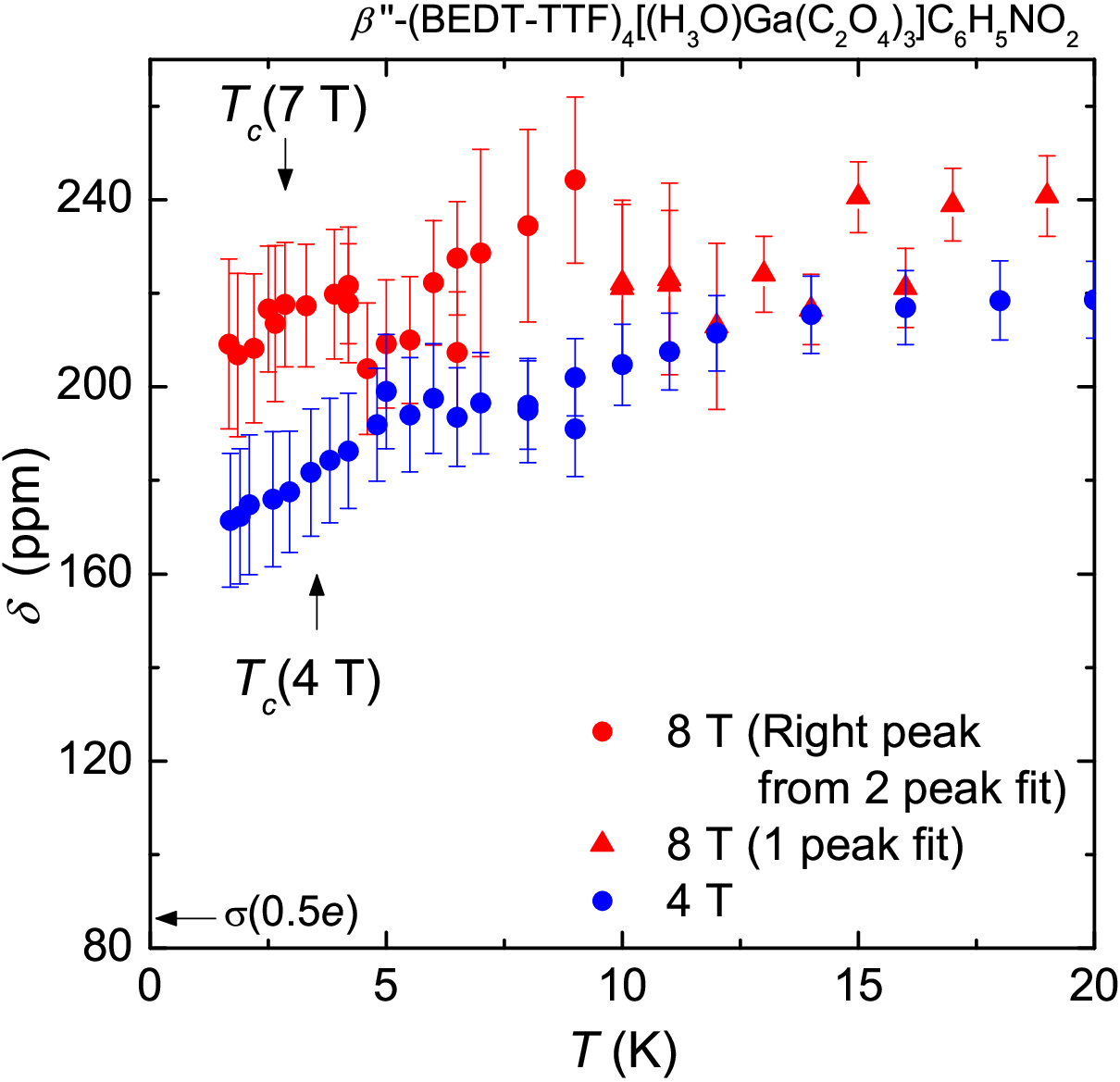}
\end{center}
\caption{(Color online)
Temperature dependence of NMR shift determined by right peak positions. 
The decrease in NMR shift below $T_{c}$ indicates a spin-singlet SC state. 
The horizontal arrow denotes the chemical shift $\sigma (0.5e)$, 
which was estimated from the $\sigma$ of $\alpha$-(BEDT-TTF)$_{2}$I$_{3}$. 
}
\label{fig4}
\end{figure}

The spin symmetry of Cooper pairs is studied by measuring the spin susceptibility $\chi_s$ in the SC state. 
We determined $\chi_s$ from the NMR shift measurement, 
as the NMR shift is the sum of the Knight shift, which is proportional to $\chi_s$, 
and the constant chemical shift $\sigma$    
($\delta = K+\sigma = A\chi _{s}+\sigma$).
The Knight shift is obtained by subtracting $\sigma$ from the NMR shift.
Because $\sigma$ is specific to the BEDT-TTF molecule and dependent only on the valence of the molecule, 
we employed a chemical-shift tensor for $\alpha$-(BEDT-TTF)$_{2}$I$_{3}$ \cite{kawai-JPSJ78} to evaluate $\sigma$ for \GaPhNo. 
For the field $H||b$, we find $\sigma(0.5e)$ to be approximately $85$ ppm. 
Figure \ref{fig4} shows the temperature dependence of the NMR shift determined by the peak position. 
Below $T_{\rm CD}$, we used right-peak positions because the NMR shift variation associated with charge disproportionation is 
small for this peak. 
At $4$ T, the reduction in NMR shift was observed below $4$ K.  
At $7$ T, only a tiny reduction was detected below $2.5$ K, 
because $T_{c}$ was suppressed by magnetic fields. 
The extrapolation of the NMR shift toward $0$ K does not reach $\sigma(0.5e) = 85$ ppm even at low fields. 
This is ascribed to the modification of $\sigma$ in the charge-disproportionate state.  
A singlet spin state is suggested for Cooper pairs, 
as $\chi_s$ decreases in the SC state. 
For a spin-singlet superconductivity, 
an unconventional SC state, such as the FFLO state, is required to account for the extremely high $H_{c2}$.  
NMR spectrum measurements in high magnetic fields are essential to unravel the SC state in high fields. 

In \GaPhNo, electron-electron correlations become strong at low temperatures 
so that the conventional Fermi liquid state is violated, and a charge-disproportionate state is realized. 
At $T_{\rm CD}$, 
we observed enhanced magnetic fluctuations possibly induced by charge fluctuations
in addition to the semiconducting resistivity.
If charge fluctuations are enhanced near $T_{c}$, 
a charge-fluctuation-induced superconductivity can be addressed for \GaPhNo, 
as theoretically predicted for $\alpha$- and $\theta$-salts \cite{merino-PRL87, merino-PRL96}. 
In $\alpha$-salts, charge instability is observed at rather high temperatures exceeding $50$ K, 
while superconductivity occurs at $T_{c}\simeq1$ K. 
Contrastingly, in $\beta''$-salts, $T_{\rm CD}$ is suppressed to $12$ K, which is close to $T_{c}$. 
Low-energy charge fluctuations may induce a high $T_{c}$ in $\beta''$-salts. 

In conclusion, the origin of NMR spectral splitting at $T_{\rm CD}$ was ascribed to a charge instability, 
because FWHM is independent of the external field.  
Although $(T_{1}T)^{-1}$ measurements at the $^{13}$C site can detect only magnetic fluctuations, 
the observed increase in $(T_{1}T)^{-1}$ is associated with charge fluctuations 
as $(T_{1}T)^{-1}$ shows a maximum at $T_{\rm CD}$. 
Below $T_{\rm CD}$, the non-Fermi liquid behavior is observed and 
superconductivity sets in at $T_{c}\simeq7$ K, where the fluctuations were still enhanced.  
We propose a relationship between charge fluctuations and superconductivity. 
The NMR shift decreases in the SC state, which is suggestive of a spin-singlet SC state.  
In order to understand the extremely high $H_{c2}$, 
an unconventional SC state, such as the FFLO state, should be taken into account.

We thank T.~Inabe for performing the X-ray diffraction experiment and H.~Kato for performing the resistivity measurement.
This work was partially supported by Grant-in-Aid for Young Scientists (B) (No. 23740249).

\bibliographystyle{apsrev}

\end{document}